\documentstyle[prl,aps]{revtex}
\begin{document}
\draft
\twocolumn[\hsize\textwidth\columnwidth\hsize\csname @twocolumnfalse\endcsname
%
%\title{SUPERCONDUCTIVITY IN THE CUPRATES AS A CONSEQUENCE
%OF ANTIFERROMAGNETISM AND A LARGE HOLE DENSITY OF STATES}
\title{ Superconductivity in the Cuprates as a Consequence
 of Antiferromagnetism and a Large Hole Density of States}
\author{E. Dagotto, A. Nazarenko, A. Moreo, and S. Haas}
\address
{Department of Physics and National High Magnetic Field Lab,
Florida State University, Tallahassee, FL 32306}
%\date{\today}
\maketitle

\begin{abstract}
We briefly review a theory for the cuprates that has been recently
proposed based on the movement and interaction of
holes in antiferromagnetic (AF) backgrounds.
A robust peak in the hole
density of states (DOS) is crucial to produce a large critical
temperature once a
source of hole attraction is identified.
The predictions of this scenario are compared with experiments. The stability
of the calculations after modifying some of the original assumptions is
addressed.
We find that if the dispersion is changed from an antiferromagnetic
band at
half-filling to a tight binding $cosk_x + cosk_y$
narrow band at $<n> = 0.87$, the main
conclusions of the approach remain basically the same i.e.
superconductivity appears in the $d_{x^2 - y^2}$-channel and $T_c$
is enhanced by a
large DOS.
 The main features distinguishing these ideas from more
standard theories based on  antiferromagnetic correlations are here discussed.

\end{abstract}
\pacs{{\bf KEY WORDS:} Superconductivity, t-J and Hubbard models, Mechanisms}
\vskip2pc]
\narrowtext

%\newpage

\noindent{{\bf 1. INTRODUCTION}}
\vskip 0.2cm

In this paper we present an informal status report of a new scenario for the
cuprates proposed
recently under the name of ``Antiferromagnetic-van
Hove'' (AFVH) theory.\cite{prl2} One of the main assumptions in this scenario
is that the normal state
of the cuprates, at least in the underdoped region, can be approximated
by a dilute gas of quasiparticles. These quasiparticles are holes heavily
dressed by spin fluctuations. The dispersion of one hole in an
antiferromagnetic background can be calculated with high accuracy using
a variety of techniques, and in the original version of
this scenario\cite{prl2} it was assumed that a
finite concentration of holes does not alter substantially
this dispersion. Of course,
it is obvious that this is an approximation and that a rigid band
filling of the hole band at hal-filling cannot be exact. But the claim is that
it is a good approximation
as the doping moves the system from an antiferromagnetic insulator
to an optimally doped superconductor.
Eventually the approximation breaks down as the AF correlations disappear.
As described below
and in a recent preprint, we have actually
shown that changing the dispersion
as the hole density increases from an AF band to a narrow tight-binding band,
as suggested by numerical simulations, does not change the
main qualitative properties originally
observed in Ref.\cite{prl2}. Thus, the rigid band
filling of the AF dispersion is $not$ a crucial assumption of the theory,
contrary to what naively can be expected.

For the interaction
between the quasiparticles a simple short range potential inspired by
calculations of two holes in the Heisenberg model is used. The
combination of these features was shown to produce a superconducting
phase in the $d_{x^2 - y^2}$-wave channel with a $T_c$ of the order
of 100K.\cite{prl2} One of the most attractive features of the
proposed scenario, that distinguishes it from other ideas based on AF
pairing,\cite{doug} is that the critical temperature acquires a
maximum at a small hole density due to the presence of a $large$ $peak$ in
the hole density of states.\cite{prl1}  It is in this respect that the present
theory has features of the more standard van Hove ideas discussed in the
literature.\cite{vanh} Details of the AFVH theory and its comparison
with experiments are given in the
following sections. For lack of space we will not be able to provide all
the many references to papers  that have contributed to the ideas described
here, and thus we apologize in advance to the readers.
All the important literature can be found from the references in the
papers quoted in this mini-review.

\vskip 0.4cm
\noindent{{\bf 2. INTERESTING TIMES IN
ANGLE RESOLVED PHOTOEMISSION EXPERIMENTS!}}
\vskip 0.2cm

Until very recently, the photoemission literature of the cuprates
contained  statements
praising the apparent good agreement between experimental data and band
structure calculations. Those results were discouraging for
experts of strongly correlated electronic models because it was clear that
the energy scales coming out from calculations in t-J models and band
structure simulations were, and still are, very different. The t-J model
near half-filling has a natural energy scale of 0.1 eV which is J, while
band structure calculations are dominated by hopping amplitudes of order eV's.
However, in recent times the
experimental data have improved dramatically and important ``surprises'' were
reported. The current picture
is quite different from what it was not long ago. Now the presence of
strong correlations is regarded as a key feature of the physics of holes
in the planes even at optimal doping where some people used to
believe that the angle-resolved photoemission (ARPES) data was
simply indicative of a gas of weakly
interacting (1-x) electrons with a dispersion derived from band
structure calculations (x is the hole concentration).

Let us here summarize the main results contained in the ARPES
data.
In this authors' opinion, there are at least four features in the recent ARPES
literature
that are clear indications of strong correlations affecting drastically
the quasiparticle dispersion:

\vskip 0.3cm

{\bf (i)} The quasiparticle bandwidth is of order J, the Cu-Cu
exchange, rather than a larger electronic scale.\cite{dessau}

{\bf (ii)} An extended region of flat ${\rm Cu
O_2}$-derived bands very near the Fermi energy exist for Bi2212,
Bi2201, Y123 and Y124.\cite{dessau}
This seems a universal property of the hole-doped cuprates.
Again, such a behavior cannot be explained within band
structure calculations which use different electronic
potentials for each compound.

{\bf (iii)} ARPES experiments for an AF $insulator$ have revealed a hole
dispersion with a small bandwidth also dominated by the exchange J.\cite{wells}

{\bf (iv)} Recent photoemission results by
Aebi et al.\cite{aebi}
in Bi2212 with
Tc=85K, using sequential
angle-scanning data acquisition to obtain PES intensities within a
narrow energy window near the Fermi energy ${\rm E_F}$,
reported evidence of antiferromagnetically induced spectral weight
above the naive Fermi momentum
${\rm {\bf p}_F}$. If
confirmed, these results tell us that there are enough AF correlations
in the normal state at optimal doping of Bi2212 to induce observable
features in a photoemission experiment. Note that several groups have
already confirmed the presence of these shadow bands in the data.
Currently the debate is whether the origin of these bands is indeed
antiferromagnetism or if it is caused by a superstructure effect.

\vskip 0.4cm
\noindent{{\bf 3. THEORETICAL PREDICTIONS
FOR THE BANDWIDTH}}
\vskip 0.2cm

Let us now compare the ARPES experiments  with predictions of
models of strongly
interacting electrons. The best model to use would be the
three band Hubbard model including degrees of freedom at both
Cu and O sites. However, this model is difficult to handle and to make
progress simplifications are necessary. Our group and many others have mostly
concentrated on the predictions arising from the 2D t-J model and from
the one band 2D Hubbard model, but we believe that the results below are
not much dependent on the specific model used
as long as there are
AF correlations in the ground state.
The ${\rm t-J}$  model Hamiltonian is defined as
\begin{eqnarray}
{\rm H  =
{}~-~t \sum_{ \langle {\bf i}{\bf j} \rangle }
({\bar c}^{\dagger}_{{\bf i}\sigma}
{\bar c}_{{\bf j}\sigma} + h.c. )
+~J \sum_{ \langle {\bf i}{\bf j} \rangle }
( {\bf S}_{{\bf i}}.{\bf S}_{{\bf j}} -
{{1}\over{4}} n_{\bf i} n_{\bf j} ),  }
\end{eqnarray}
\noindent
where the notation is standard. The one hole dispersion in this model
has been calculated in recent years by a variety of techniques
both analytical and
numerical. Recently, some of us in collaboration with M. Boninsegni
studied this dispersion,
$\epsilon ( {\bf k} )$, using a
Green Function Monte Carlo (GFMC) method\cite{prl1} on
large clusters of $8\times8$,
$12\times 12$ and $16\times16$ sites minimizing finite
size effects. Other techniques produce similar results.\cite{review}
In Fig.1a, the numerically evaluated  $\epsilon ( {\bf k} )$
is shown at ${\rm J/t=0.4}$ along particular directions in the
Brillouin zone.
The minimum in the energy is obtained at the ${\rm {\bar M}}$ point
${\bf k} = (\pi/2, \pi/2)$ in agreement with several
previous approximate
calculations.
The main $effective$ contribution to $\epsilon ( {\bf k})$ arises from
hole hopping between sites belonging to
the same sublattice, to avoid distorting the AF background. The density
of states (DOS) is given in Fig.1b.
The total bandwidth, ${\rm W}$, is severely reduced from that of a gas
of non-interacting electrons
due to the antiferromagnetic correlations
(the vertical axis in Fig.1a,b is in units of the hopping t,
which is ${\rm \approx 0.4 eV}$). Many features of this dispersion are
 in nice agreement
with the experimental result (i) described above.\cite{prl1}
Note also that all momenta
belonging to the non-interacting 2D Fermi surface
${ cosk_x + cosk_y = 0}$ are
very close in energy in the t-J model.
In the range ${\rm 0.3 \leq J/t \leq 0.7}$, we found that
the energy difference between ${\rm (\pi/2,\pi/2)}$ and ${\rm (0,\pi)}$
is approximately 15\% to 20\% of the total bandwidth. The proximity of
all the states along the ${ cosk_x + cosk_y = 0}$ line contributes to a large
peak in the density of states which is a crucial feature of the theory
for the cuprates  discussed below.
Note also that
since ${\rm W \sim J}$, and more importantly since the splitting between
points along the $X=(\pi,0)$ to $Y=(0,\pi)$ line is even smaller than J,
 then a non-negligible temperature dependence is
expected in the data.
Such small energy scales in the problem produce a Hall
coefficient ${\rm R_H}$
in quantitative agreement with the experimental results for LSCO. \cite{prl1}

It is also interesting
to notice the evolution of the Fermi surface from hole pockets to a
large FS as the hole doping increases in reasonable agreement with
experiments.\cite{prl1} These results were also noticed by S. Trugman
time ago.\cite{trugman}
For details the reader should consult Ref.\cite{prl1} and references
therein.
Eder and Ohta have also done important contributions to these ideas in recent
papers.\cite{eder}
It is interesting to remark that mean-field spin-density-wave
calculations in the literature for the one band Hubbard model do $not$
remove the degeneracy along the $(\pi,0)$ to $(0,\pi)$ model, affecting
severely the results. The removal of this degeneracy is very important
to properly describe the physics of the model.

\vskip 0.4cm
\noindent{{\bf 4. FLAT REGIONS IN THE CUPRATES}}
\vskip 0.2cm

An interesting detail of Fig.1a is the near flatness of the energy
in the vicinity
of ${\bf k} = (\pi,0)$. This feature is similar to
the ARPES results described in (ii) above.
Actually, in Refs.\cite{prl2,prl1} the
theoretically observed flat region of Fig.1a was explicitly
compared with ARPES experiments finding a good agreement.
The flatness of the ${\bf k } = (\pi,0)$
region in the cuprates has a many-body origin, rather than being
induced by band effects (see also Refs.\cite{bulut1}).
Intuitively, it is the presence
of saddle-points near X and Y (also noticed in Ref.\cite{trugman}) plus
the presence of a near degeneracy along $(\pi,0)$-$(0,\pi)$
which is responsible for the abnormal flatness
of the hole dispersion. These combined effects produce
a density of states
DOS with a robust peak at the bottom of the hole band (top of the
valence electronic band) and  a van-Hove (vH) singularity (Fig.1b).
This large DOS was used by some of us to boost the critical temperature
once a pairing mechanism was identified.\cite{prl2} As remarked before,
the presence of this
van Hove singularity and the large DOS has strong similarities with vH
scenarios discussed before in the literature.\cite{vanh}
In the previous vH ideas the dispersion
is usually generated by band effects, while
here $many-body$ effects
in the ${\rm CuO_2}$ planes are crucial, and thus our predictions are
universal for all hole-doped cuprates.

\vskip 0.4cm
\noindent{{\bf 5. ARPES DATA FOR A HOLE IN AN ANTIFERROMAGNETIC INSULATOR}}
\vskip 0.2cm

In this section, we will discuss the interesting results
obtained by Wells et al.\cite{wells} for the
AF insulator ${\rm Sr_2 Cu O_2 Cl_2}$.
The main
conclusion of the experimental effort is that the theory for one
hole in the 2D t-J model describe very well the data  in almost all
directions in momentum space. In particular, the agreement along the
$(0,0)$ to $(\pi,\pi)$ line is remarkable. This result
is encouraging since it shows that indeed the 2D t-J model
is suitable for a first principles description of the
${\rm CuO_2}$ planes. However, there are discrepancies in the vicinity of the
$(0,\pi)$  and $(\pi,0)$ points. The prediction of the t-J model would
indicate a flat energy close to the top of the valence band as described
before. However, the ARPES result has a larger binding energy.\cite{wells}
In the opinion of the authors, we have to keep in perspective that this
discrepancy cannot be labeled as a failure of the model since it is
clear to all of us that the t-J model is only a rough approximation and
the hope is that it captures some of the essential features of the experiments.
Asking for a perfect agreement with the data is too much. With this
caveat, however let us also say that the absence of the flat regions
near the $(0,\pi)$ and $(\pi,0)$ is an issue to worry about since they
are a very clear feature of the other cuprates at optimal doping. Then, an
important question arises: is the absence of flat regions in
${\rm Sr_2 Cu O_2 Cl_2}$ an indication that this material is different
from the rest? After all ${\rm Sr_2 Cu O_2 Cl_2}$ is not a
superconductor since it is difficult to dope.
A possibility raised by some of us is that small
modifications around the t-J model may account for the ARPES dispersion
of this compound. The modification amounts to the addition of a
hopping amplitude for next to nearest neighbor sites.
This modified model describes better the data than the t-J
model.\cite{gooding} However, recent preliminary
ARPES results by the Stanford group
for underdoped Bi2212 suggest that as a function of doping, moving from
the AF insulator to optimal doping, the region from $(0,0)$ to $(\pi,\pi)$
remain the same while the neighborhood of $(0,\pi)$,$(\pi,0)$ is much
affected.\cite{priv} The flat regions seem to emerge
as the doping grows towards
its optimal value. Work is in progress to analyze this crucial aspect of the
new ARPES data. Results will be presented soon.

\vskip 0.4cm
\noindent{{\bf 6. SHADOW BANDS}}
\vskip 0.2cm

As described in Sec.2, recent experiments by Aebi et al. have generated
considerable excitement.\cite{aebi} A feature
presumably caused by antiferromagnetic
correlations was reported in their ARPES data for Bi2212.
 This result is compatible with the
 ``shadow bands'' scenario of Kampf
and Schrieffer\cite{spinbag}  which is a consequence of
antiferromagnetic
correlations in the normal state.
At half-filling, these bands, which appear at momenta
above the naive Fermi momentum ${\rm {\bf p}_F}$,
are caused by the enlarged magnetic unit cell of the
${\rm CuO_2}$ planes produced by the long range antiferromagnetic
order in the ground state.
This effective reduction in the size of the Brillouin
zone (BZ) has interesting implications for
PES experiments.\cite{spinbag} For example, along the
diagonal ${\rm p_x = p_y = p}$, and assuming long-range order,
peaks at momenta ${\rm {\bf p_1}= (p,p)}$
and ${\rm {\bf p_2} = (\pi - p, \pi - p)}$
should appear at the same energy location,
for any value of ${\rm p}$. The PES weight ($\omega < 0$) observed in
the region $above$ the non-interacting ${\rm {\bf p}_F }$
is induced by strong magnetic correlations.\cite{spinbag}
How important is this antiferromagnetically generated PES weight
at finite density? Only recently
Quantum Monte Carlo and Exact Diagonalization calculations in the proper
regime of strong coupling have
been discussed.\cite{haas}. At density $\langle n \rangle = 0.88$ and
coupling ${\rm J/t=0.4}$, the AF correlation is of only two lattice
spacings as in optimal YBCO. In spite of this small correlation, a sharp
peak in the hole spectral function
was observed in Ref.\cite{haas} near the chemical potential even
for momenta ${\rm (2\pi/3,2\pi/3)}$ i.e. above the naive Fermi momentum.
This peak is correlated with AF and disappears with further increasing
of the density. But at $\langle n \rangle = 0.88$ it is still
observable, giving more support to the AF interpretation of Aebi et al.'s
data\cite{aebi}
and to theoretical scenarios based on
AF.\cite{spinbag,bickers,prl2} In short, an
antiferromagnetic
correlation of two lattice spacings can produce observable results in
PES experiments for the cuprates! This remarkable result was unexpected
since the previous intuitive perception
was that very short range correlations are
mostly irrelevant. Our results show that this is
incorrect and, specially for the dispersion of holes, the spin ordering
in its immediate vicinity is important. If the ``shadow bands'' found in
Bi2212
turn out to have a superstructure explanation rather than an AF origin,
then the numerical simulation results can be used to conclude
that the AF correlation
of Bi2212 should  be smaller than a couple of lattice spacings,
making it negligible. Either way, the study of shadow bands in the
cuprates is important.

\vskip 0.4cm
\noindent{{\bf 7. d-WAVE SUPERCONDUCTIVITY}}
\vskip 0.2cm

The overall conclusion of the
 previous sections is that the quasiparticles described by
the 2D t-J model are in many respects in good agreement with ARPES data.
Let us now introduce interactions among the
quasiparticles. We will follow ideas based on
antiferromagnetism to produce the pairing attraction
needed for superconductivity. However, there is
an important distinction with respect to previous
``AF-oriented'' literature\cite{bickers}: in the AFVH scenario
the DOS of the quasiparticles
has a large peak that in a natural way induces the
existence of an optimal doping i.e. a density at which
the critical temperature is maximized.

To build up a model for the cuprates we
construct the interaction between the quasiparticles
based on the 2D t-J model, where
it is well-known that in an antiferromagnet an
effective attractive force exists between two holes leading to a bound
state in the ${\rm d_{x^2 - y^2}}$-wave channel.\cite{review}
As shown numerically in many studies, the dominant effective
attraction is between nearest-neighbors sites. Thus, the
model we will use in our analysis is
\begin{equation}
{\rm H =-\sum_{{\bf p},\alpha} \epsilon_{\bf p}
c^{\dagger}_{{\bf p}\alpha} c_{{\bf p}\alpha}
- |V| \sum_{\langle {\bf ij} \rangle} n_{\bf i} n_{\bf j} },
\end{equation}
where ${\rm c_{{\bf p}\alpha}}$ is an operator that destroys a
quasiparticle with momentum ${\rm {\bf p}}$ in sublattice
$\alpha = {\rm A,B}$;
${\rm n_{\bf i}}$ is the number operator at site ${\bf i}$; ${\rm |V|
= 0.6J}$ (which can be deduced from the ${\rm t-J}$ model\cite{prl2}),
and $\epsilon_{\bf p}$ the dispersion evaluated
in Ref.\cite{prl1}. No double occupancy is allowed. Since in
the original t-J language quasiparticles
with spin-up(down) move in sublattice A(B), the interaction term can
also be  written as a spin-spin interaction.
This Hamiltonian has been constructed
based on strong AF correlations, and it has a vH singularity and a large
peak in the ``noninteracting'' hole
DOS. We will refer to it as the AFVH model.

We studied the AFVH model with the standard BCS formalism.
Since ${\rm |V|/W \sim 0.3}$, where ${\rm W}$ is the bandwidth of the
quasiparticles, the gap equation should produce a reliable
estimation of Tc since we are effectively
exploring the ``weak'' coupling regime of the AFVH model.
Solving the gap equation on $200 \times 200$ grids,
we observed that the free energy is minimized using a ${\rm
d_{x^2-y^2}}$ order parameter.
In Fig.2, ${\rm T_c}$ against
the hole density is shown.

Two features need to be
remarked: i) an optimal doping exists at which ${\rm T_c}$ is maximized
which is a direct consequence of the large peak in the DOS
of the quasiparticles. Note that such a peak would
have an important effect even in theories where the hole interaction is
phonon mediated rather than spin-wave mediated. In other words, one of
the main
concepts introduced in Ref.\cite{prl1} and \cite{prl2}, i.e.
a large peak in the DOS generated by antiferromagnetism, is obviously not
restricted to models where
superconductivity is produced by an electronic mechanism;
ii) the optimal doping (15\%) and
optimal ${\rm T_c \sim 100K}$  are in good agreement with
the cuprates phenomenology. Although in the AFVH model the
natural scale of the problem is ${\rm J \sim 1000K}$,
since the ratio
between coupling and bandwidth is small, ${\rm T_c}$ is further reduced
in the weak coupling BCS formalism to about 100K.
Note that this quantitative agreement with experiments is
obtained without the need of ad-hoc fitting parameters.

Note also that the AFVH model Eq.(2) seems
to play for d-wave the role that the attractive Hubbard model plays
for s-wave superconductivity. In particular, the two body problem leads to
a d-wave bound state, a feature that other phenomenological models of
d-wave superconductivity do not have.\cite{micnas} This issue has been
discussed in a recent publication where exact diagonalization results on
a 32 site lattice have shown the existence of d-wave superconductivity
in the AFVH model analyzed beyond the BCS gap equation.\cite{newmodel}

The ratio ${\rm R(T)}$
${\rm = 2 \Delta_{max}({\rm T})/{\rm k T_c} }$ can be calculated from
the gap equation,
(for a $d_{x^2-y^2}$-wave condensate, ${\rm \Delta_{max}({\rm T}) }$ is defined
as the maximum value of the gap).
At ${\rm T=0}$, the AFVH model predicts ${\rm R(0) =
5.2}$ while recent
tunneling experiments\cite{gap} give 6.2.
Other experiments have reported a smaller
value for ${\rm R(0)}$. For example, ARPES data by Ma et al.\cite{ma}
obtained
${\rm R(0) = 4.6}$, while an average over
the pre-1992 literature\cite{batlogg} suggested
${\rm R(0) = 5 \pm 1}$ supporting the results of the
AFVH model.
We have also verified that an important
feature of previous vH scenarios\cite{vanh}
also exists in our model, i.e. a quasiparticle lifetime linear with
frequency at the optimal doping\cite{prl2}.

A standard concern of any scenario that makes use of a van Hove
singularity is its stability after the addition of disorder that tend to
rapidly smear the logarithmic singularity and reduce drastically its
effects. Also in models where the van Hove singularity appears in the
noninteracting limit, the presence of strong correlations tend to
destroy such a singularity. In our AFVH scenario none of these effects
are important. Note that strong correlations have already been taken into
account in the construction of the quasiparticle dispersion and the
expectation is that these q.p.'s are weakly interacting. Note also that the
van Hove singularity of the DOS at half-filling appears on top of a
large accumulation of weight naturally caused by features of a hole in
an antiferromagnet as described before. Then, although the actual
log-singularity may disappear with disorder, the robust peak in the DOS
remains, as was shown explicitly in Ref.\cite{prl2}. Then, this model does not
have the weaknesses of other van Hove based models in the literature.

\vskip 0.4cm
\noindent{\bf 8. EFFECT OF A FINITE HOLE DENSITY ON THE HOLE DISPERSION}
\vskip 0.2cm

An immediate concern about the AFVH scenario is the issue
of the influence of hole density over the quasiparticle dispersion. In
the construction of Sec.7. it was assumed that holes still move as they
were moving in a N\'eel state even with other holes present.
It is obvious that this assumption cannot be
exact i.e. as the density diminishes from half-filling changes must
occur in the hole dispersion since the AF correlation length diminishes
as the number of holes grow. In Ref.\cite{finite} this issue was addressed
numerically using the exact diagonalization technique on a finite
lattice. There, working at
$<n>=0.87$  the result shown in Fig.3
was found. The PES part of the spectrum is very similar to that found
at half-filling giving support to the assumptions of Sec.7. The near
flat region seems still present in the PES data, although now the error
bars and finite size  effects are more severe than they were at
half-filling.
On the other hand, note
that there is a substantial IPES region that contributes appreciably to
the dispersion (see also Ref.\cite{bulut1}). In Fig.4 we show that
the hole DOS at finite density still has a large peak at the top
of the band as at half-filling. This peak is crossed
by the chemical potential as the doping is moved
away from half-filling.\cite{stanford,plakida}

To study whether our scenario will be modified
by the IPES contribution, we carried out the following exercise:
instead of using the AFVH dispersion in the kinetic energy of Eq.(2), we
used a $cosk_x + cosk_y$
dispersion to mimic the broad features of
Fig.3 with the amplitude $t_{eff}$ as a
fitting parameter ($t_{eff}$ is still much smaller than the bare hopping $t$).
Using this dispersion,
i.e. entirely neglecting the ``AF-induced'' part of the spectrum
observed in Fig.3 we carried out the calculation of the critical
temperature using the same interaction as before i.e. a nearest
neighbor attraction of order J. As shown by inspection in Fig.3
the chemical potential
in this quasiparticle band is near the flat region and in the language
of the tight-binding ``cos+cos'' dispersion it amounts to working with a
half-filled q.p. band. Then,
effectively the superconductivity
that will arise after hole attraction is introduced will
correspond to that of a narrow band ``t-U-V model''
at half-filling which we know leads to a
d-wave condensate.\cite{micnas} The new
$T_c$ is shown in Fig.5. Although much reduced with respect to the
original AFVH result\cite{prl2},
it still has a robust value of $\sim 35K$ since the
density of states is again
large due to the approximate flat regions near the
saddle points of the tight-binding
dispersion. In Ref.\cite{dos} we have also introduced a dispersion that
interpolates between the two extreme cases (AFVH and tight binding). The
$T_c$ in this case is also shown in Fig.4 and it reaches $\sim 60K$.
Then, the main quantitative conclusions
of the AFVH scenario remain unaltered i.e. even using a
dispersion obtained numerically and removing the AF induced shadow part,
 the robust peak
in the DOS still produces a robust $T_c$, and the channel remains
$d_{x^2 - y^2}$-wave. Details are
presented in a recent preprint.\cite{dos} These results show that our scenario
is robust under reasonable modifications of the main assumptions. Such
stability is important. Our ideas are general and do not correspond
to a fine tuning of parameters to fit the data.

We have also done the following exercise: using
the AFVH dispersion, we added to
the one particle Green functions that enter
into the gap equation a factor $Z_{\bf
k}$ in the numerator to account for the fact that the quasiparticles do
not carry 100\% of the weight in $A({\bf k},\omega)$. Actually to study
a drastic influence of $Z_{\bf k}$, we made zero the
weight of those states which were at a distance larger than
20 meV from the flat
regions. This trick can also account for the
fact that the quasiparticle peaks away from the Fermi surface become broad
with a width growing like the square of the energy distance from the
Fermi surface, as in any Fermi
liquid. The results, recently discussed in Ref.\cite{dos}, are given in
Fig.5. The main features of the original AFVH result
remain qualitatively the same after these modifications are
introduced. It is becoming clear that the
main factor contributing to the large  $T_c$ are the
flat regions at
the top of the valence band where most of the spectral weight is
accumulated. The rest of the dispersion, and with it most of the
AF-induced shadow band, is not as important. Then, as it happened before for
the case of disorder and strong correlations, our approach seems
stable against changes in the quasiparticle dispersion to account
for the finite hole density.

\vskip 0.4cm
\noindent{\bf 9. PHASE SEPARATION:}
\vskip 0.2cm

It has been known since the early studies of the t-J model that as
J/t grows at any finite hole
density, the system will eventually phase  separate.
Near half-filling this is caused by the attraction between holes coming
from  the minimization of broken antiferromagnetic links.\cite{review} The
presence of a tendency to form large domains of holes has been detected
in several high-Tc
materials in their normal state.\cite{hammel,domains} The philosophy of
our group is that  phase separation mechanisms and AF-based mechanisms
correspond to different sides of the same coin. In the general phase
diagram of the t-J model it is well-known that, at a fixed density, as
J/t grows we move from a metallic phase, to a superconducting phase, to
a phase separated regime.\cite{review}
The force leading to superconductivity and
phase separation is basically the same in this model. The region of
major interest, i.e. the superconducting regime, can be accessed from the
paramagnetic region by developing AF correlations or from the phase
separated region by reducing these correlations, i.e. improving the
hole mobility.\cite{emery} Then, to understand the superconducting phase both
approaches are supposed to lead to the same result.

\vskip 0.4cm
\noindent{\bf 10. IS RETARDATION IMPORTANT? CAN PHONONS MATTER AFTER ALL?}
\vskip 0.2cm

A feature of theories of high-Tc which is rarely addressed is the
importance of ``retardation'' effects. In the old superconductors
described by a phononic mechanism these effects are clearly crucial since the
delay in the interaction introduced by the exchange of phonons (with a
typical velocity related to Debye energies rather than to Fermi
energies) manages to avoid the instantaneous Coulombic repulsion
between the two electrons forming a Cooper pair. In theories of
high-Tc, somehow
we should also be able to avoid the repulsion between holes in the pair.
However, an immediate problem arises related to the fact that the
spin-wave velocity is governed by the exchange J, while near
half-filling the bandwidth of the hole band is
also of order J, as shown before. Thus, the Fermi velocity is not
obviously larger than the spin-wave velocity i.e.
there is no clear distinction between the velocities of the (rather
heavy) dressed quasiparticles and the bosons mediating the attraction.
 In order to avoid the Coulombic repulsion one may
use the following argument: a
pair of holes in an antiferromagnet have a natural finite distance between the
members of the pair since the attractive channel is d-wave. The pair
cannot be ``on-site''. However,
this is not enough to avoid entirely the problem since at a distance of one
lattice spacing the naive bare Coulomb repulsion is large enough
to destroy the weak
attraction used in our AFVH Hamiltonian. A possible way to avoid this
problem is by invoking a large polarization (i.e. a large dielectric
constant) or a strong screening effect. But a more interesting possible
explanation is that the hole attraction in the t-J model indeed has some
important but somewhat hidden retardation effects. Suppose that a hole in an AF
produces ``string'' states i.e. once the hole is injected at a given
site its movement leads to the creation of a string of spins that are not
correctly aligned with respect to the staggered background.\cite{review}
This effect costs
energy and the only way to avoid it is by the hole to retrace its path, unless
quantum spin fluctuations destroy the string. Let us imagine what occurs
in the small J/t limit. Here a hole can move a large distance since
the energy paid by the string creation grows like J, while the mobility
is regulated by t. But eventually the hole feels the string which
makes it return to the origin.
But
a second hole added to the problem
would improve its energy by taking advantage of this string, i.e.
moving along the path opened by the first hole is energetically
favorable since the string is erased. Then, we can roughly envision a
hole pair as
one hole that leads the way creating a string and a second hole that
follows through exactly the same path, healing the damage. The
two holes remain at some prudential distance to avoid the Coulombic
repulsion. We believe that this picture
may lead to an interesting retardation effect that may solve
the Coulomb repulsion problem in AF scenarios for the cuprates.

There is another way to avoid the Coulombic interaction that we
have recently studied. It may occur that the strength of the AF
correlations in the normal state at optimal doping is enough to
affect the quasiparticle dispersion, as described before, but it may
not be strong enough to produce pairing. Then, a model where the holes
approximately have the dispersion arising from AF, but with
the pairing mediated
by phonons would be a possibility. The main trouble that immediately
arises is that the natural channel for phononic mechanisms is s-wave.
However, we found that the buckling mode of YBCO can produce attraction
in the d-channel.\cite{phonons} This fact was also observed by Song and
Annett.\cite{song} One of the clear advantages of this approach is that
a nonzero isotope effect can be explained in a natural way. Current
experimental information indeed favors the presence of a nonzero isotope
coefficient away from optimal doping. Although
the idea of mixing AF concepts with phonons is
not in such a developed state as those based on AF mediated
pairing, the possibility of phonons for d-wave deserves further study.

\vskip 0.4cm
\noindent{\bf 11. CONCLUSIONS}
\vskip 0.2cm

Recent ARPES experimental results have provided evidence that strong
correlations are important to properly describe the physics of the
quasiparticle carriers in the
cuprates even at optimal densities. We attribute the features observed in
ARPES data to strong AF correlations and based on them, we have built a
theory of high-Tc combining van Hove and AF scenarios. Concrete
predictions come out from our approach
namely $d_{x^2 - y^2}$-wave superconductivity,
and the concept of optimal doping when the chemical potential reaches
the large peak in the DOS. The results are not much affected if the
rigid band filling of the hole dispersion at half-filling is relaxed
by using a more standard tight-binding quasiparticle dispersion as given by
numerical studies of the 2D t-J model. The theory is also robust after
the introduction of disorder, and it produces a Hall coefficient which
agrees well with experiments.\cite{prl1}  A long paper with the details
of our calculations will be presented soon.

\vskip 0.4cm
\noindent{\bf ACKNOWLEDGMENTS}
\vskip 0.2cm

Discussions with many colleagues have helped in the development
of these ideas. The list is too long to be given here. We are grateful to
all of them. E.D. and A.M. are supported by the Office of Naval Research under
grant ONR N00014-93-0495. Additional support by Martech and the NHMFL is
also acknowledged.

{\bf Figure Captions}

\begin{enumerate}

\item (a) Energy of a hole
in the ${\rm t-J}$ model, $\epsilon ( {\bf
k})$, vs momentum obtained with the GFMC method on a $12 \times 12$
lattice (open squares) and ${\rm J/t = 0.4}$ (in units of t). Results
for an $8\times8$ cluster (open triangles)
and a $16\times16$ cluster (full squares) are
shown.
(b) Density of states obtained from our fit of the numerical
data Fig.1a
showing the van-Hove singularity between ${\rm {\bar M}}$ and X.
The unit of energy is ${\rm t}$ (from Ref.\cite{prl1}).

\item Critical
temperature ${\rm T_c}$ of the AFVH model as a function of hole
density ${\rm x}$ ($=1-\langle n \rangle)$ (using the BCS gap
equation). The superconducting state is
$d_{x^2-y^2}$-wave (from Ref.\cite{prl2}).

\item Dispersion of the hole quasiparticle in the 2D t-J model at
J/t=0.4 and density $<n>=0.87$. The chemical potential is at zero.
$\omega >0$ is the IPES region. The size of the open and full
circles is proportional to the intensity of the quasiparticle peaks.
The technique used is exact diagonalization. For details see
Ref.\cite{finite}

\item Density of states of the 2D t-J model at finite electronic
density, and J/t=0.4. The results were obtained with exact
diagonalization. For details see Ref.\cite{stanford}.

\item Critical temperature using the AF original dispersion used in
Ref.\cite{prl2} (dashed line), using the tight-binding dispersion
described in Sec.8 and Ref.\cite{dos} (dot-dashed line) and also using
a combination of both as in Ref.\cite{dos} (solid line).

\end{enumerate}

\end{document}